\documentclass{ws-ijmpa}

\begin{document}

\markboth{J.~Przerwa et al. -- COSY--11 Collaboration}
{J.~Przerwa et al. -- COSY--11 Collaboration}

%
\catchline{}{}{}{}{}
%

\title{ISOSPIN DEPENDENCE OF THE $\eta^{\prime}$ MESON PRODUCTION
IN NUCLEON-NUCLEON COLLISIONS}

\author{\footnotesize J.~Przerwa$^{\star}$$^,$\footnote{E-mail address: 
j.przerwa@fz-juelich.de}~, 
P.~Moskal$^{\star,\$}$,
A.~Budzanowski$^{\#}$, 
E.~Czerwi{\'n}ski$^{\star,\$}$,
R.~Czy{\.z}ykiewicz$^{\star}$,
D.~Gil$^{\star}$,
D.~Grzonka$^{\$}$,
L.~Jarczyk$^{\star}$,
T.~Johansson$^{@}$,
B.~Kamys$^{\star}$,
A.~Khoukaz$^{\%}$,
P.~Klaja$^{\star}$,
W.~Krzemie{\'n}$^{\star,\$}$,
W.~Oelert$^{\$}$,
C.~Piskor-Ignatowicz$^{\star}$,
J.~Ritman$^{\$}$,
B.~Rejdych$^{\star}$,
T.~Sefzick$^{\$}$,
M.~Siemaszko$^{\&}$,
M.~Silarski$^{\star}$,
J.~Smyrski$^{\star}$,
A.~T{\"a}schner$^{\%}$,
M.~Wolke$^{\$}$,
P.~W{\"u}stner$^{\$}$,
J.~Zdebik$^{\star}$,
M.~J.~Zieli{\'n}ski$^{\star}$,
W.~Zipper$^{\&}$
}

\address{
$^{\star}$Institute of Physics, Jagellonian University, Cracow, Poland\\ 
$^{\$}$IKP and ZEL, Forschungszentrum J{\"u}lich,J{\"u}lich, Germany\\ 
$^{\#}$Institute of Nuclear Physics, Cracow, Poland\\
$^{@}$Department of Physics and Astronomy, Uppsala University, Sweden\\
$^{\%}$IKP, Westf{\"a}lische Wilhelms-Universit{\"a}t, M{\"u}nster, Germany\\
$^{\&}$Institute of Physics, University of Silesia, Katowice, Poland\\ 
}

\maketitle


\begin{abstract}
A comparison of the close--to--threshold total cross section for the $\eta'$
production in  $pp~\to~pp\eta'$ and $pn~\to~pn\eta'$ reactions constitutes
a tool to investigate the $\eta'$ meson structure and the reaction mechanism 
in the channels of isospin I~=~0 and I~=~1 and
may provide insight into the flavour-singlet
(perhaps also into gluonium) content of the $\eta'$ meson.\\ 
In this contribution we present preliminary results 
of measurement 
of the quasi-free 
production of the $\eta'$ meson in
the proton-neutron collisions
conducted using the COSY-11 facility.

\end{abstract}

\section{Introduction}  
In the framework of the quark model the $\eta^{\prime}$ meson is predominantly
a flavour-singlet combination of quark-antiquark pairs, and it is expected
to mix with purely gluonic states.
Therefore, additionally to the production mechanisms associated with meson
exchange\cite{nakayama01,kampfer} 
it is also possible that $\eta^{\prime}$ meson is produced from excited glue
in the interaction region of the colliding nucleons, which couple to the $\eta^{\prime}$
meson directly via its gluonic component or through its SU(3)-flavour-singlet
admixture\cite{bass02}. As suggested in reference\cite{bass03},
the $\eta^{\prime}$ meson production via the colour-singlet object does not depend on the
total isospin of the colliding nucleons and should lead to the same amplitudes of the production
for the $pn \to pn\eta^{\prime}$ and $pp \to pp\eta^{\prime}$ reactions.\\
In case of the $\eta$ meson, the ratio of the total cross sections for the reactions
$pn \to pn\eta$ and $pp \to pp\eta$ was determined to be $R_{\eta}$~$\approx~$~6.5\cite{calen01},
and $R_{\eta}$~$\approx$~3 at threshold\cite{pneta},
what suggest the dominance of isovector meson exchange in the $\eta$ production in nucleon-nucleon
collisions.
Since the quark structure of $\eta$ and $\eta'$ mesons is similar,
in case of the dominant isovector meson exchange -- by the analogy to
the $\eta$ meson production -- we can expect that the ratio $R_{\eta'}$
should be large. If however $\eta'$ meson is produced via its
flavour-blind gluonium component from the colour-singlet glue excited
in the interaction region the ratio should approach unity after corrections
for the initial and final state interactions.
The close--to--threshold
excitation function for the $pp \to pp\eta'$
reaction has already been established\cite{ppetap}
and the determination of  the total cross section for the $\eta'$ meson production in the proton-neutron
interaction constitutes the aim of the work reported in this contribution.

\section{Experiment}
\begin{figure}[h]
\parbox{0.45\textwidth}{\psfig{file=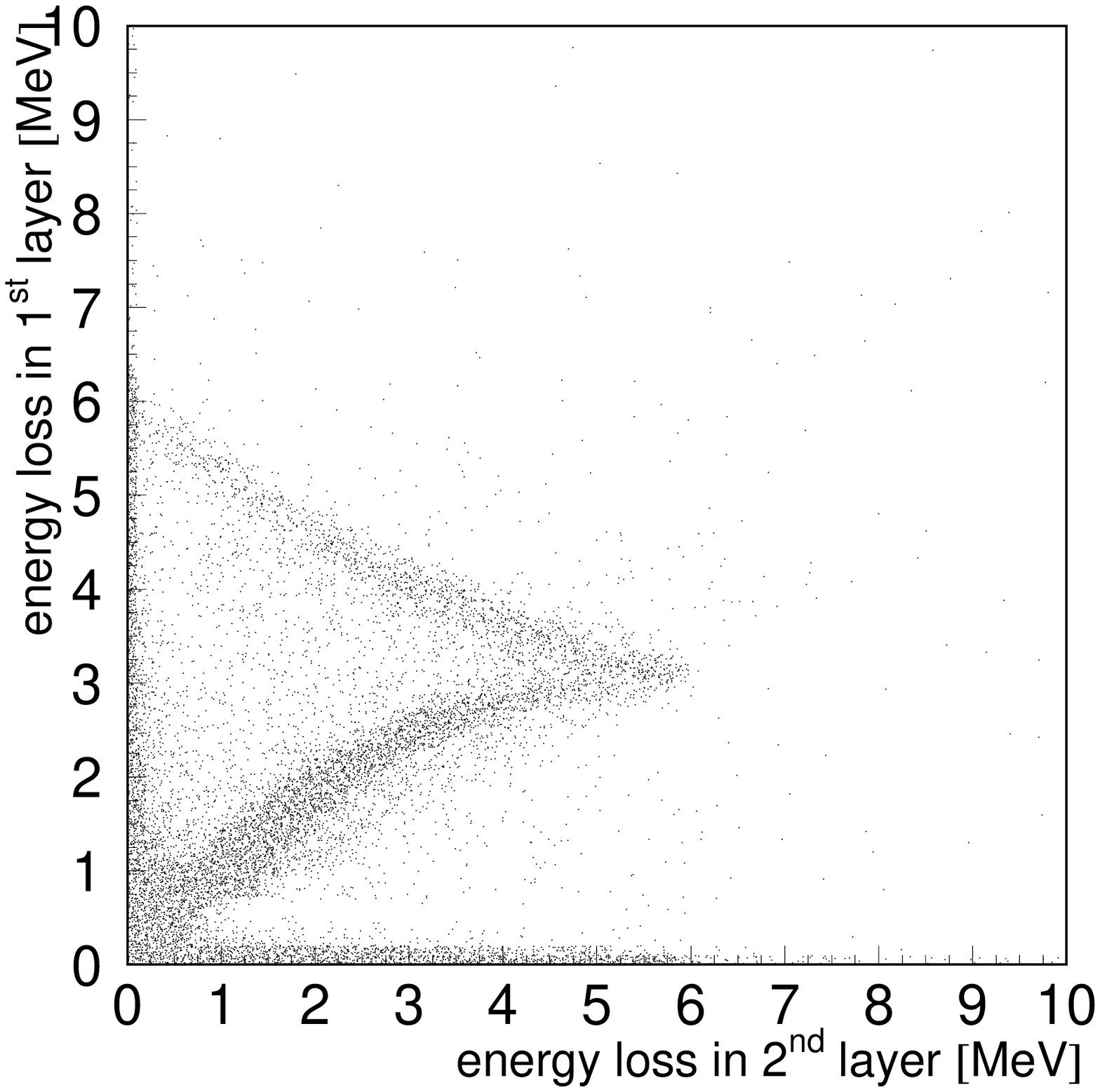,width=6.5cm}} \hfill
\parbox{0.45\textwidth}{\psfig{file=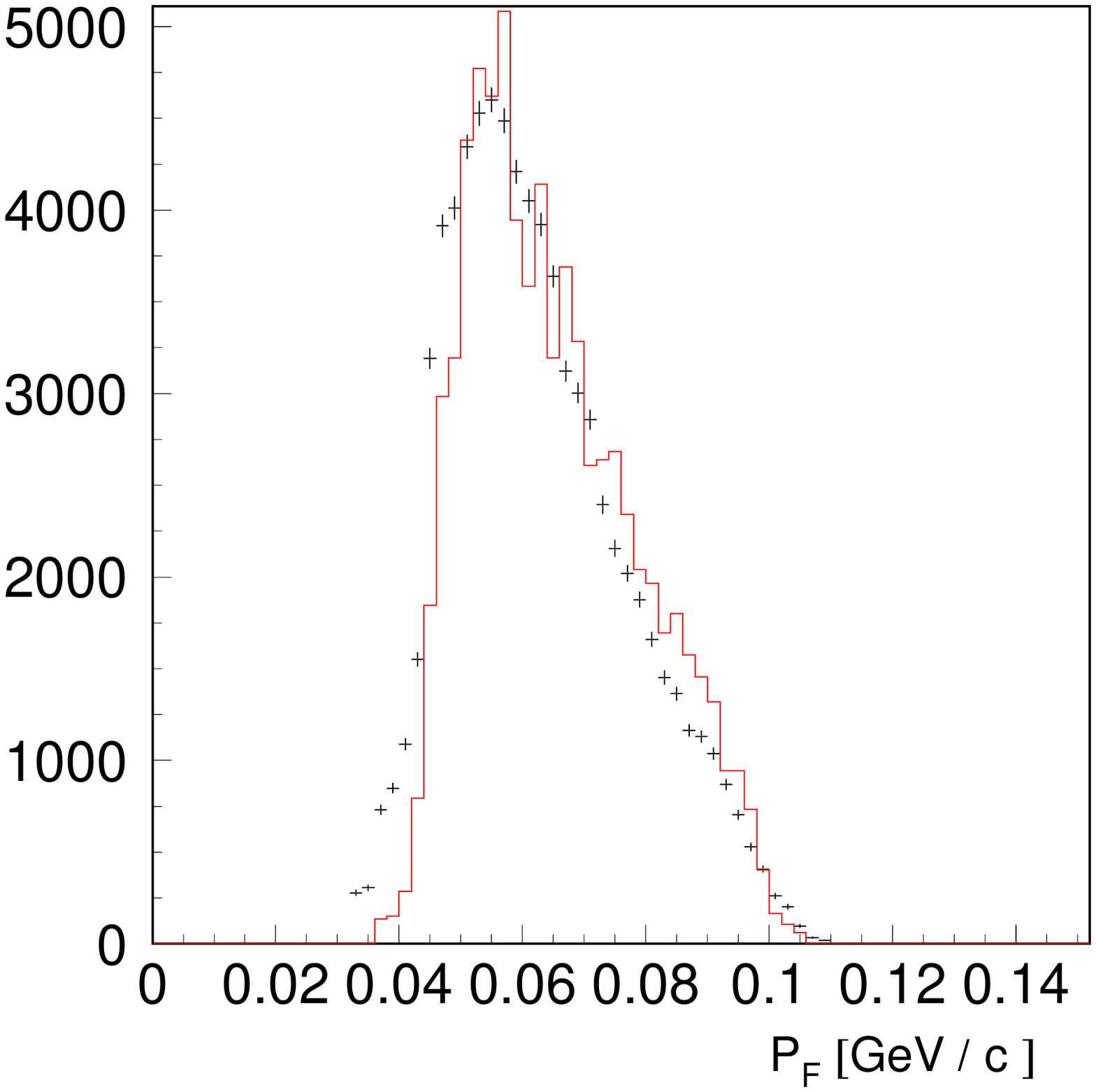,width=6.5cm}}
\caption{{\bf Left}: Energy losses in the first layer versus 
                     the second layer as measured at COSY--11 with
                     a deuteron target and a proton beam with momentum
                     of 3.35 GeV/c. 
        {\bf Right}: Momentum distribution of the proton spectator
                     as reconstructed in the experiment (points)
                     in comparison with simulation taking into account  Fermi 
                     momentum distribution of nucleons inside the deuteron (solid histogram).}      
\label{fig1}
\end{figure}

In August 2004 using the COSY--11 facility\cite{cosy11}
we have conducted a measurement of the $\eta'$ meson production
in the proton-neutron collision.
The experiment has been realized using a proton beam of the cooler synchrotron COSY~\cite{cosy}
and a cluster jet deuteron target~\cite{dombrowski}.  
Proton and neutron
outgoing from the $pn\to pn\eta$ reaction have been registered by means
of the COSY-11
facility.
For the data analysis
the proton from the deuteron is considered as a spectator which does not interact
with the bombarding proton, but escapes untouched and hits the detector
carrying the Fermi momentum possessed exactly at the time of the reaction.
The experiment is based on the registration of all outgoing nucleons
from the $pd\to p_{sp} pnX$ reaction~\cite{pneta}.
Fast protons are measured in two drift
chambers and scintillator detectors~\cite{cosy11}, neutrons are registered
in the neutral particle detector~\cite{przerwa01}, and
slow spectator protons moving upwords to the beam are measured by the dedicated
silicon-pad detector~\cite{bilger01}. Fig.~\ref{fig1}(left)
shows energy losses in the $1^{st}$ layer of the spectator detector versus
$2^{nd}$ layer. 
Slow spectator protons are stopped
in the first or second layer of the 
detector whereas 
fast particles cross both detection layers.
Having the deposited energy and the emission angle  we calculate
the kinetic energy of the spectator proton and its momentum.
Fig.~\ref{fig1}(right) shows momentum distribution
of protons considered as a spectator as determined at COSY-11
with a deuteron target and a proton beam with momentum of 3.35~GeV/c
(points) compared with simulations taking into account a Fermi motion of nucleons inside the deuteron (solid line).
 
\begin{figure}[h]
\parbox{0.45\textwidth}{\psfig{file=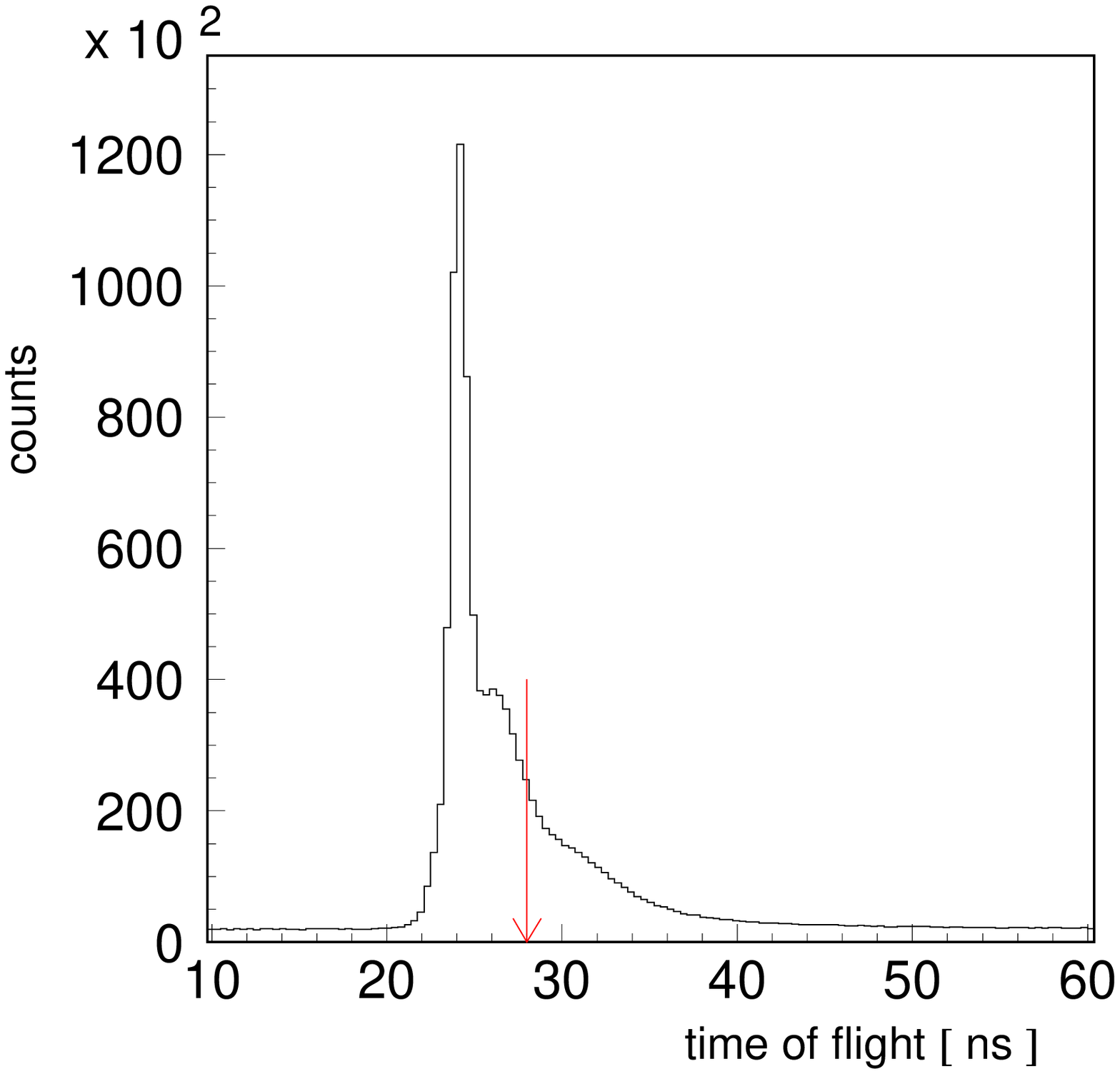,width=6.7cm}} \hfill
\parbox{0.45\textwidth}{\psfig{file=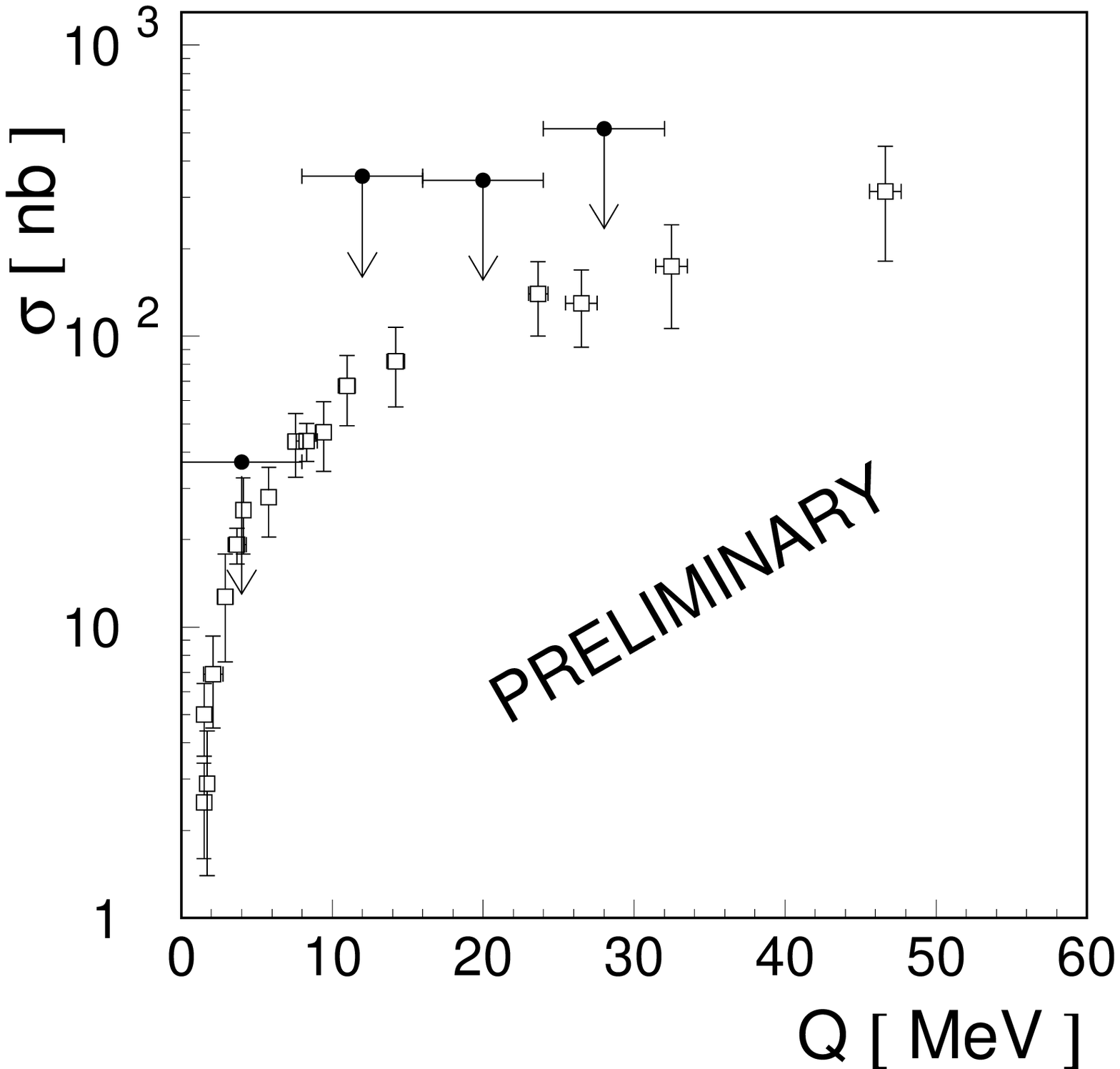,width=6.7cm}}
\caption{{\bf Left}: Time--of--flight distribution between the target
                     and the neutron detector as obtained after the time
                     walk correction under condition that in coincidence
                     with neutral particle also two charged particles
                     were identified. 
        {\bf Right}: Total cross sections for the $pp \to pp\eta^{\prime}$
                     reaction as a function of the excess energy (open symbols).
                     Upper limit for the total cross section for the $pn \to pn\eta^{\prime}$
                     reaction as a function of the excess energy (closed symbols).}
\label{fig2}
\end{figure}

Application of the missing mass technique allows to identify events
with the creation of the meson under investigation.
The total energy available for the quasi-free proton-neutron
reaction can be calculated for each event from the vector of the momenta
of the spectator and beam protons.
The absolute momentum of neutrons is determined from the time-of-flight between
the target and the neutron detector.
Fig.\ref{fig2}(left) presents the time-of-flight distribution -- for
neutral particles -- measured between the target and the neutral particle
detector.
A clear signal originating from the gamma rays is seen
over a broad enhancement from neutrons. This histogram  shows that discrimination
between signals originating from neutrons and gamma quanta can be done by a cut
on the time of flight. From the Monte Carlo simulations  
of the $pn \to pn\eta^{\prime}$ reaction
the largest expected momentum value of the neutron is eqaul to 1.4~GeV/c
which corresponds to the time--of-flight value of 28.5~ns 
as it is indicated by an arrow in Fig.~\ref{fig2}(left).
 Neutrons
having time--of--flight below this value originate from $pn \to pn~pions$ 
reactions and are not taken for the further analysis.\\
\vspace{-1cm}
\vspace{-1cm}

Due to the smaller efficiency and lower resolution for the registration
of the quasi-free $pn \to pn~meson$
reaction in comparison to the measurements of the proton-proton reactions,
the elaboration of the data encounters problems of low statistics.
Therefore, the excess energy range for $Q \ge 0$ has been divided
only into four intervals of 8~MeV width. For each interval
we have calculated the missing mass. Next, from events with negative Q value 
the  corresponding background missing mass
spectrum was constructed, shifted to the kinematical limit and normalized to the experimental distribution
at the very low mass values where no events from the $\eta^{\prime}$ are expected.  
Detailed description of the method used for the background subtraction can be found in the dedicated article\cite{pmoskal}.
After subtracting missing mass distributions for the negative values of Q from spectra for Q values
larger than 0 -- 
due to the very high signal--to-background ratio -- 
at the present stage of the data analysis,
the signal from the $\eta^{\prime}$ meson was found to be statistically insignificant.
Nevertheless, having the luminosity  --~established from 
the number of the quasi-free
proton-proton elastic scattering events~\cite{aiplumi}~--
and the detection efficiency of the COSY-11 system we have estimated
the upper limit of the total cross section for the quasi-free 
$pn \to pn \eta^{\prime}$ reaction. The prelimnary result is shown
in Fig.~\ref{fig2}(right).

\section{Acknowledgments}
We acknowledge the support by the
European Community
under the FP6 programme (Hadron Physics,
RII3-CT-2004-506078), by
the Polish Ministry of Science and Higher Education under grants
No. 3240/H03/2006/31, 1202/DFG/2007/03, 0082/B/H03/2008/34,
and by the German Research Foundation (DFG).


\begin{thebibliography}{9}
\bibitem{nakayama01}
K.~Nakayama et al., {\it Phys. Rev.} {\bf C 61}, 024001 (2000).
\bibitem{kampfer}
L.~P.~Kaptari, B.~K{\"a}mpfer, e-Print: arXiv:0804.2019.
\bibitem{bass02}
S.~D.~Bass, {\it Phys. Scripta} {\bf T 99}, 96 (2002).
\bibitem{bass03}
S.~D.~Bass, {\it Phys. Lett.} {\bf B 463}, 286 (1999).
\bibitem{calen01}
H.~Calen {\it et al.}, {\it Phys. Rev.} {\bf C 58}, 2667  (1998).
\bibitem{pneta} 
P.~Moskal {\it et al.}, e-Print: arXiv:0807.0722 (2008).
\bibitem{ppetap}
P.~Moskal {\it et al.}, {\it Phys. Lett.} {\bf B 474}, 416 (2000);\\
A.~Khoukaz {\it et al.}, {\it Eur. Phys. J.} {\bf A 20}, 345 (2004);\\
F.~Balestra {\it et al.}, {\it Phys. Lett.} {\bf B 491}, 29 (2000);\\
P.~Moskal {\it et al.}, {\it Phys. Rev. Lett.} {\bf 80}, 3202 (1998).
\bibitem{cosy11}
S.~Brauksiepe {\it et al.}, 
{\it Nucl. Instr. $\&$ Meth.} {\bf A 376}, 397 (1996);\\
P.~Klaja {\it et al.}, 
{\it AIP Conf. Proc.} {\bf 796}, 160 (2005);\\
J.~Smyrski {\it et al.}, 
{\it Nucl. Instr. $\&$ Meth.} {\bf A 541}, 574 (2005).
\bibitem{cosy} 
D.~Prasuhn {\it et al.}, 
{\it Nucl. Instr. \&  Meth.}  {\bf A 441}, 167 (2000).
\bibitem{dombrowski} 
H.~Dombrowski {\it et al.}, 
{\it Nucl. Instr. \& Meth.} {\bf A~386}, 228  (1997).
\bibitem{przerwa01}
J.~Przerwa {\it et al.}, 
{\it Int. J. of Mod. Phys.} {\bf A 20}, 625 (2005).
\bibitem{bilger01}
R.~Bilger {\it et al.}, 
{\it Nucl. Instr. $\&$ Meth.} {\bf A 457}, 64 (2001).
\bibitem{pmoskal}
P.~Moskal {\it et al.}, 
{\it J. Phys.} {\bf G 32}, 629 (2006).
\bibitem{aiplumi} 
P. Moskal, R. Czy{\.z}ykiewicz, 
\emph{AIP Conf. Proc.} {\bf 950}, 118 (2007).

\end{thebibliography}
\end{document}